# Direct observation of electron density reconstruction at the metal-insulator transition in NaOsO$_3$

N. Gurung,[1,2] N. Leo,[1,2] S. P. Collins,[3] G. Nisbet,[3] G. Smolentsev[2], M. García-Fernández,[3] K. Yamaura,[4] L. J. Heyderman,[1,2] U. Staub[2], Y. Joly[5], D. D. Khalyavin[6], S. W. Lovesey[3,6] and V. Scagnoli[*,1,2]

[1]*Laboratory for Mesoscopic Systems, Department of Materials, ETH Zurich, Switzerland*
[2]*Paul Scherrer Institute, Villigen PSI, Switzerland*
[3]*Diamond Light Source Ltd, Didcot, Oxfordshire OX11 0DE, United Kingdom*
[4]*Research Center for Functional Materials, National Institute for Materials Science, 1-1 Namiki, Tsukuba, Ibaraki 305-0044, Japan*
[5]*Univ. Grenoble Alpes, CNRS, Grenoble INP, Institut Néel, 38000 Grenoble, France*
[6]*ISIS Facility, STFC Oxfordshire OX11 0QX, UK*

5$d$ transition metal oxides offer new opportunities to test our understanding of the interplay of correlation effects and spin-orbit interactions in materials in the absence of a single dominant interaction. The subtle balance between solid-state interactions can result in new mechanisms that minimize the interaction energy, and in material properties of potential use for applications. We focus here on the 5$d$ transition metal oxide NaOsO$_3$, a strong candidate for the realization of a magnetically driven transition from a metallic to an insulating state exploiting the so-called Slater mechanism. Experimental results are derived from non-resonant and resonant x-ray single crystal diffraction at the Os L-edges. A change in the crystallographic symmetry does not accompany the metal-insulator transition in the Slater mechanism and, indeed, we find no evidence of such a change in NaOsO$_3$. An equally important experimental observation is the emergence of the (300) Bragg peak in the resonant condition with the onset of magnetic order. The intensity of this space-group forbidden Bragg peak continuously increases with decreasing temperature in line with the square of intensity observed for an allowed magnetic Bragg peak. Our main experimental results, the absence of crystal symmetry breaking and the emergence of a space-group forbidden Bragg peak with developing magnetic order, support the use of the Slater mechanism to interpret the metal-insulator transition in NaOsO$_3$. We successfully describe our experimental results with simulations of the electronic structure and with an atomic model based on the established symmetry of the crystal and magnetic structure.

**Keywords:** Metal-insulator transition, x-ray resonant scattering, magnetism, Slater insulator

# I. INTRODUCTION

The metal-insulator transition, MIT, has been a key topic of condensed matter physics since Verwey's pioneering work on magnetite, the prototype of this class of transition [1, 2]. The strong interest arises from its deep links to the fundamental interactions of correlated electron physics, and poses a major challenge in our understanding of complex systems. Until now, $3d$ transition metal oxides have been widely studied displaying striking phenomena including high temperature superconductivity (most notably in cuprates [3] and pnictides [4, 5]), colossal magnetoresistance [6, 7], and metal-insulator transitions [1, 8]. This variety of phenomena stems from the several competing interaction terms associated with the charge, orbital and magnetic degrees of freedom.

Recent studies prove that $5d$ transition metal compounds are just as fascinating as they display several striking physical properties. Due to the larger spatial extent of the $5d$ orbitals in comparison to $3d$ orbitals, $5d$ transition metal compounds experience a large crystal field splitting of $t_{2g}$ and $e_g$ states ($\approx$ 2 eV above the Fermi energy) and thus, notably, have relatively weak electronic correlations according to conventional understanding. While the electron correlation, parametrized by the Hubbard correction term $U$, diminishes in size on descending the periodic table from $3d$ to $5d$ elements, the spin-orbit interaction increases in value with increasing atomic number. In addition, there is a strong $p$-$d$ hybridization resulting from the large orthorhombic distortion that is caused by the octahedral rotation, which affects the bandwidth in these simple perovskites. Orthorhombic distortion here refers to the distortion from ideal cubic to orthorhombic structure (e.g. tilting and rotation of $OsO_6$ octahedra). To give an estimate of such distortions, it is customary to introduce the Goldschmidt tolerance factor [9], t, where t = 1 for ideal cubic perovskite structure. The calculated value for $NaOsO_3$ is t=0.9077 and similar perovskite compounds have: t($NdNiO_3$)= 0.913; t($GdFeO_3$)= 0.9021; t($TbMnO_3$)= 0.8505. In recent studies, focus was mainly on the interplay between the spin-orbit coupling, the Coulomb repulsion and the bandwidth in Iridates [10-13] and to a lesser extent in Osmates [14-20]. Such a flurry of research in $5d$ systems has resulted in the discovery of topological insulating phases, spin liquid behavior [21] and bulk insulating states [10].

Particularly intriguing is the nature of the insulating state in these $5d$ systems. It is currently under debate as to whether this is of either Mott [11, 22, 23] or Slater [15-19] character. The latter mechanism proposed by Slater more than 50 years ago [24] has the emergence of antiferromagnetic ordering as the source of an electron localization that drives the system into an insulating state. This insulating state is very distinct from the metallic phase that characterizes the paramagnetic temperature regime above the Neel ordering temperature. In such a model, electron correlations do not play any role, which is in contrast to the case of the Mott scenario where it is the electron correlations that favor the existence of an insulating ground state. However, no clear evidence of a compound supporting a magnetically driven metal-insulator "Slater" transition has been found to date. Among the most likely candidates that might exhibit a Slater metal-insulator transition, the most promising is the perovskite $NaOsO_3$ [16, 17, 19].

$NaOsO_3$ undergoes a metal-insulator transition and antiferromagnetic ordering at the same temperature, $T_{MIT}= T_N \sim 410$ K. This is a remarkably high temperature for a compound for which the Coulomb electronic correlations should be weaker than, or at most, comparable to the energy scale of the spin-orbit interaction. The magnetic moment determined by neutron refinement is 1 $\mu_B$ [16], which is less than expected from the nominal valence $Os^{5+}$ with 3 electrons singly occupying the $t_{2g}$ levels as predicted from Hund's rules. A reduced moment is suggestive of the itinerant nature of the $5d$ electrons and of significant hybridization with neighboring oxygen orbitals. The $NaOsO_3$ crystallographic and magnetic structures have been studied and a G-type antiferromagnet with magnetic moments lying along the *c*-axis (*Pnma* space group) below $T_N$ was reported as well as the presence of a small ferromagnetic moment along the *b*-axis [15, 16]. The magnetic space-group was determined to be *Pn'ma'* (#62.448), which belongs to the centrosymmetric crystal class *m'mm'*. Small crystallographic changes suggested [16] the absence of a crystallographic phase transition in the vicinity of $T_N \sim 410$ K, but there is also an observed anomaly in the *a* and *c* lattice constants in the vicinity of $T_{MIT}= T_N$. The metal-insulator transition in $NaOsO_3$ is known to be a second order phase transition from a metallic (above 410 K) to an insulating state [15, 16]. The concomitant MIT and antiferromagnetic ordering, and possible absence of crystallographic symmetry breaking, are suggestive of a Slater MIT. However, due to the presence of energetically similar competing interactions, a consensus on the nature of the metal-insulator mechanism operating in this perovskite is absent. Current literature on $NaOsO_3$ includes pressing arguments for three mechanisms, namely, a spin-driven Lifshitz mechanism using a magnetic reconstruction of the Fermi surface [25], the aforementioned Slater mechanism [16, 17], and a Mott-Hubbard mechanism that is independent of magnetic correlations and is result of an electron localization effect controlled by an on-site strong Coulomb interaction that overcomes the delocalization, which is determined by measuring the bandwidth [20]. In light of its intriguing properties,

NaOsO$_3$ has been the subject of several experimental [15, 16, 19, 26] and theoretical [17, 18, 25] studies.

In order to validate the Slater mechanism, it is important to demonstrate the absence of crystallographic symmetry change occurring across the metal-insulator transition. Evidence from x-ray and neutron powder diffraction suggests that this crystallographic symmetry change is absent. Although, powder diffraction methods are sensitive to lattice distortions, they are lesser sensitive to symmetry breaking with only smooth variations of lattice constants. In recent years, resonant x-ray scattering has proven to be a powerful technique to detect symmetry breaking by measuring weak intensities at forbidden reflections [27] that are directly related to the symmetry of the charge density of the resonant atoms. This sensitivity arises from the fact that, at resonance, the x-ray scattering is strongly enhanced and phenomena that are usually negligible, such as the asphericity of the atomic electron density, can be observed. Thus, at resonance, the x-ray scattering factor is no longer a scalar and must be treated as an anisotropic tensor (creating an "anisotropy of the tensor of scattering" (ATS) or Templeton-Templeton scattering) [28]. As a result, tuning the incident x-ray energy, e.g. to the Osmium L$_3$ edge, gives high sensitivity to the Osmium's coordination with its nearest neighbors, as well as to spatial distortions and anisotropies of the Osmium electron density. Resonant scattering thus provides the possibility to make a quantitative study of the microscopic mechanism causing the MIT by observing the modification of the Os electron density occurring across the MIT.

Henceforth, we have conducted resonant x-ray scattering experiments at the Os L edges on a small single crystal of NaOsO$_3$ to determine whether or not there is a change in the crystallographic symmetry across MIT. Also, we have focused on the forbidden reflections, (300) and (030), and compared the experimental results with simulations performed with the FDMNES package [29] as well as with atomic model calculations, which provide direct insight into the physical nature of the observed scattering intensity.

This paper is organized as follows: in Section II and III we describe the sample preparation and the x-ray resonant scattering technique used to perform the measurements. In Section IV, we present the experimental results that provide evidence of a change in the diffracted intensity of the (300) forbidden reflection across the metal-insulator transition. In Section V, we compare our observation with the FDMNES simulations and with an atomic model calculation, from which we extract quantitative information on the nature of the change in the electronic structure across the phase transition.

## II. EXPERIMENTAL DETAILS

Single crystal samples of NaOsO$_3$ were grown in pressures up to 6 GPa as described in Ref. [15]. Several single crystals with sizes of approximately 0.2 mm x 0.3 mm x 0.1 mm were oriented with x-ray Laue back reflection and then polished mechanically to have a well-defined surface perpendicular to either the [100] or the [010] direction. Further x-ray characterization with Cu $K\alpha$ radiation in an x-ray diffractometer enabled the selection of high quality crystals, having small mosaicity, resulting in rocking curves with several Bragg reflections e.g. (400), (600), (220) with full width half maximum of ~ 0.01°.

Resonant x-ray scattering experiments were carried out in the vicinity of the Osmium L$_3$ and L$_2$ edges at beamline I16 at the Diamond Light Source. The beamline is equipped with a 6-kappa diffractometer that can operate in horizontal and vertical scattering geometries, with the scattered x-rays in the plane and perpendicular to the plane of the electron storage ring respectively. The beamline has a double-bounce silicon harmonic rejection mirror system that provides exceptionally high harmonic rejection over a wide energy range. The beamline is equipped with a Silicon monochromator with an energy resolution of ~1.5 eV at the energies corresponding to the Osmium L edges. The incident radiation was linearly polarized perpendicular to the vertical scattering plane (σ polarization) with a beam size of 0.2 mm (horizontal) × 0.03 mm (vertical). A graphite (008) crystal was used at the Os L$_3$ edge for polarization analysis of the diffracted beam (with the state of polarization denoted by primed quantities). For σ − π′ scattering, the suppression of the σ − σ′ channel was approximately 99.9%, and vice versa. A Pilatus 100 K photon-counting pixel detector was used for the measurements performed without polarization analysis. The Pilatus detector pixel size (0.172 mm x 0.172 mm) results in an estimated momentum resolution of 0.0016 Å$^{-1}$ and 0.0004 Å$^{-1}$ for measurements performed with incident photon energy of 10.785 keV and 5.2 keV, respectively. Throughout the paper we use the symbol $\psi$ to designate the azimuthal angle, which represents a rotation of the sample around a selected diffraction wave vector. The azimuthal angle reference position was chosen to be zero when the [100] and the [010] directions are in the scattering plane.

Diffraction and absorption signals across the Osmium L edges (such as those illustrated in Fig.1) are simultaneously recorded from two different regions of interest with the Pilatus pixel detector, while measuring the energy dependence of the intensity of the diffraction peaks. Sharp multiple scattering contributions to the diffracted intensity were determined and minimized by performing several energy dependence scans of the same reflection, with slightly different azimuthal angles.

We complemented this characterization with absorption measurements in transmission geometry at the SuperXAS beamline at the Swiss light source in the vicinity of the Osmium L edges and crystallographic single crystal diffraction performed with x-rays having an incident energy of 16 keV at the Swiss-Norwegian beamline at the ESRF [30].

### III. RESONANT X-RAY SCATTERING

Resonant x-ray scattering combines the chemical sensitivity of absorption (spectroscopy) and the atomic position sensitivity of diffraction. The measured intensity is proportional to the square of the unit cell structure factor. To maximize the sensitivity of the technique to detect crystallographic symmetry breaking, measurements are typically focused on forbidden or weakly allowed reflections. For such diffraction conditions, only the scattering originating from the resonant ions (weighted by a phase factor which depends on the atom position and the scattering wave vector) contributes to the diffracted intensity. To correctly compute the diffraction intensity, all contributions to the x-ray atomic scattering factor $f$ should be considered. The most general expression for $f$ is:

$$f = f_o + f_m + f' + if''$$

where $f_o$ corresponds to the classical Thomson scattering of the atom and $f_m$ is the non-resonant magnetic scattering amplitude. $f'$ and $f''$ are the two energy-dependent anomalous dispersion correction terms of the atomic scattering factor. In particular, $f'$ and $f''$ describe the resonant scattering amplitude arising from photon assisted electronic transitions between core and empty states, with a scattered photon subsequently re-emitted when the electron and the core hole recombine [31]. When the photon energy approaches the energy corresponding to an absorption edge, the spectroscopy becomes sensitive to the unoccupied states just above the Fermi level. These unoccupied states are highly sensitive to the local environment with its symmetry and possible distortion (see e.g. Fig. 6a), and provide an indirect probe of the corresponding electronic density. This means that, at resonance, the scattering can be strongly modified and phenomena that can be usually neglected, such as the asphericity of the atomic electron density, can be determined [28, 32, 33]. By determining the dependence of the Bragg peak intensity on the x-ray energy, polarization, and azimuthal angle rotation, one can distinguish between structural (non-resonant scattering from the $f_o$ term), electronic (e.g. "charge ordering" or "charge disproportionation" and ATS) and spin (magnetic scattering) contributions, depending on the chosen reflection (which is simply related to the Fourier components of the specific long-range order under investigation). Such sensitivity is due to the fact that, at resonance, the x-ray scattering factor is no longer a scalar quantity and should be treated as an anisotropic tensor [34]. The tensorial nature of the scattering process offers the possibility to directly determine tiny changes in the electronic (or magnetic) structure of the sample with incomparable sensitivity with respect to other techniques such as neutron diffraction that are not directly sensitive to electronic ordering.

Following Ref. [35, 36], in the most general case the structure factor, $F$ for a given reflection of index (*hkl*) can be described as:

$$F_{hkl} = \sum_{K,Q,q}(-1)^Q H^K_{-Q} D^K_{Qq} \Psi^K_q \qquad (1)$$

The positive integer $K$ is the rank of the multipole, and the projection $q$ can take $(2K + 1)$ integer values that satisfy the relationship $-K \leq q \leq K$. The first term $H^K_{-Q}$ describes the dependence of the structure factor on the polarization of the incoming and outgoing x-rays, $D^K_{Qq}$ reflects a rotation of the local axes of the sample required to fulfill the selected diffraction condition (depending on the orientation of the sample compared to the surface normal) and $\Psi^K_q$ is given by:

$$\Psi^K_q = \sum_d e^{i\mathbf{d}\cdot\mathbf{\tau}} \langle T^K_q \rangle_d \qquad (2)$$

where $\langle T^K_q \rangle_d$ is an atomic multipole that represents the electronic origin of the scattering, with the index $d$ labeling the position of the resonant ion in the unit cell. $\mathbf{\tau} = (hkl)$ is the scattering wave-vector. Angular brackets $\langle \ldots \rangle$ denote the expectation value of the enclosed electronic spherical tensor operator, which is defined in reference [35]. For an electric-dipolar (E1-E1) transition, multipoles of rank $K$ up to 2, contribute to the absorption cross section. The $0^{th}$ rank term ($K = 0$) corresponds to an isotropic atomic charge contribution given for example by the presence in the sample of resonant atoms with different non integer valence (e.g. $Ni^{3+\delta}$ and $Ni^{3-\delta}$, with $\delta$ being slightly different form zero [27]). Such a charge imbalance is commonly referred to as "charge disproportionation". The $K = 1$ terms correspond to time-odd dipole contributions (e.g. a magnetic dipole), and $K = 2$ corresponds to time-even quadrupoles (which reflect, for example, the degree of hybridization with neighboring ions, e.g. ATS).

The structure factors for specific Bragg reflections in $NaOsO_3$ and their angular dependence are derived in Appendix C.

## IV. RESULTS

For the Slater model of a metal-insulator transition, significant lattice distortions are not required to accompany the emergence of the magnetic ordering and the appearance of the insulating state. Therefore, the first step to validate the Slater model is to confirm the absence of crystallographic symmetry breaking at the metal-insulator transition. A common strategy here is to verify the absence of any diffraction intensity at positions in reciprocal space corresponding to forbidden reflections for the crystallographic space group of the sample. A crystallographic phase transition, such as the one from orthorhombic to monoclinic symmetry, such as that observed in some of the perovskite nickelates [27, 37-39], results in the loss of some space group specific symmetry elements. This changes the selection rules for forbidden reflections, leading to the emergence of a measurable intensity at specific reciprocal lattice points. $NaOsO_3$ has already been investigated extensively by x-ray and neutron powder diffraction [15, 16]. However, for very tiny deformations, these methods might not be sensitive enough to detect a symmetry change in the sample, if not guided by some compelling evidence from other experimental results suggesting such a symmetry break. Therefore, in order to test whether a crystallographic symmetry break occurs at the metal-insulator transition of $NaOsO_3$, we took advantage of the superior sensitivity of single crystal diffraction over powder diffraction methods. We performed a systematic search below the Neel temperature $T_N$ for a diffraction signal at positions in reciprocal space corresponding to forbidden reflections, using an x-ray energy of 5.2 keV, which is far from the Osmium absorption edges. This choice of low x-ray energy also minimizes intensities arising from "parasitic" signals due to multiple scattering of x-rays within the sample due to the limited number of reflections in the Ewald sphere. No intensity, from either crystallographic symmetry breaking or non-resonant magnetic scattering was observed in a reciprocal lattice scan with an acquisition time of 20 seconds per point. This indicates the absence of crystallographic symmetry breaking and non-resonant magnetic scattering, with $NaOsO_3$ remaining in the same *Pnma* space group above and below the metal-insulator transition. Such results were subsequently corroborated by x-ray single crystal structural determination on smaller crystallites, performed at the SNBL beamline at the ESRF.

Having confirmed that no crystallographic symmetry breaking takes place across the metal-insulator transition in $NaOsO_3$, we then focused our attention on the observation of possible changes in the electron density occurring at the metal-insulator transition. Such a variation in the electron density would provide insight into whether or not a specific electron localization pattern occurs, e.g. charge ordering or orbital ordering, as observed in manganites [40-48] or nickelates [27, 39]. Such an observation would yield new insight into the nature of the metal-insulator transition and shed light upon the validity and, if relevant, on the nature of the Slater mechanism in NaOsO$_3$.

Therefore, we have performed resonant diffraction measurements at x-ray photon energies corresponding to the $L_3$ and $L_2$ Osmium edges ($2p \rightarrow 5d$ electronic dipolar transitions). Resonant x-ray diffraction was used to study selected magnetic and ATS (sometimes also called "orbital ordering") diffraction peaks in NaOsO$_3$. The possibility to tune the energy of the incident x-rays to the Osmium $L$ edges provides element-specific electronic density distribution information that is not directly accessible with any other technique. By selecting a specific reflection, we are able to observe a specific component of the distorted charge density.

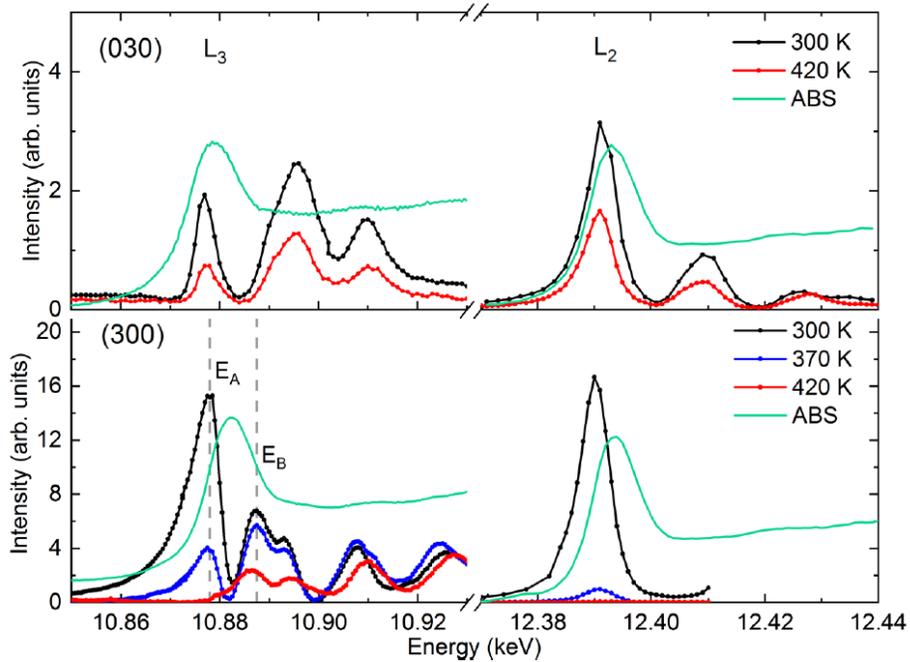

FIG. 1. *(Color online) (Top panel) Energy dependence of the intensity of the (030) forbidden reflection at fixed momentum transfer for selected temperatures illustrating the peak intensity variation across the MIT at Os $L_3$ (left) and $L_2$ edge (right) edges for $\psi \sim 90°$. The x-ray absorption spectra (ABS) measured at the same energy range are also shown. Spectra are not corrected for absorption. (Bottom panel) Same as above, but for the (300) reflection. The azimuthal angle in this case was $\psi \sim 0°$.*

For the specific case of a $d^3$ ion, and in the absence of hybridization, we expect a relatively weak intensity for the ATS peaks as the singly occupied $t_{2g}$ orbitals form a representation of an angular momentum of $L = 1$ with zero total angular momentum [18], which implies almost cubic symmetry. In our experiments we have chosen to focus our attention on the (030) and (300) ATS reflections. At such reflections, the strong magnetic contribution (in the magnetically ordered phase), which would make the measurements of the ATS contribution challenging, is absent. Both reflections were studied above and below the metal-insulator transition ($T_N$ = $T_{MIT}$ = 410 K). The variation of the intensity of such reflections as a function of the x-ray energy, in the vicinity of the $L_3$ and $L_2$ edges is given in Fig. 1. By comparing the energy spectra in the metallic (> 410 K) and the insulating phase (< 410 K), we have found an important difference between the two (030) and (300) ATS reflections. While the (030) reflection displays only a gradual change of intensity as function of temperature, with no change in shape, the spectrum of the (300) undergoes a dramatic change across the metal-insulator transition at $T_N$~411K. At both the $L_3$ and $L_2$ edges, an extra feature appears in the spectra in the insulating state (for the $L_3$ edge the relevant feature is indicated by $E_A$ in Fig. 1) corresponding to the inflection point of the absorption spectrum. This energy typically corresponds to an electronic resonance involving the unoccupied $5d$ orbitals, and to changes that could correspond to the emergence of either magnetic ordering or electron structure rearrangement.

To determine the origin of the change in the spectral shape of the (300) reflection at the energy $E_A$= 10.878 keV, we performed a characterization of the variation of its intensity as a function of the azimuthal angle $\psi$ and also as a function of the polarization of the scattered x-rays which was then compared with the structure factor expression given by Eq. 1. The azimuthal angle dependence presented in Fig. 2, was recorded without a polarization analyzer since preliminary measurements with polarization analysis at selected azimuthal angles showed intensity only in the rotated $\sigma - \pi'$ polarization. The fact that there is no intensity in the $\sigma - \sigma'$ channel indicates that "charge ordering" or "charge disproportionation" (associated with the $K$=0 term in Eq. (1)) cannot contribute to the observed (300) diffracted intensity. "Charge ordering" requires the presence of Os atoms in different oxidation states, whose scattering would interfere constructively and give a finite intensity in the $\sigma - \sigma'$ channel. The fact that the Os atoms have the same oxidation state is consistent with the reported crystallographic structure.

Assuming the reported magnetic structure [16] is correct, no magnetic signal (contributions from rank $K = 1$ tensors) should contribute to the (300) reflection as defined by the structure factor in Eq. 1. Therefore, we consider only the contribution to the diffracted intensity related to the electric quadrupole $K = 2$ term. Applying symmetry arguments as described in the Appendix, one obtains the following expression for the expected azimuthal angle dependence for the different polarization channels of this reflection: $F_{\sigma\sigma'} = F_{\pi\pi'} = 0$ and $F_{\sigma\pi'} \propto \langle T_1^2 \rangle' \cos(\theta) \cos(\psi)$, where $\theta$ is the Bragg angle, $\psi$ is the azimuthal angle and $\langle T_1^2 \rangle'$ is the real part of the tensor $\langle T_1^2 \rangle$.

The observed azimuthal intensity (Fig. 2) displays within experimental uncertainties, a cosine squared modulation, which is consistent with the modulation expected from the space group symmetry, *Pnma,* given that the origin of the signal is ATS (anisotropic scattering).

However, a magnetic origin for the (300) reflection at $E_A$ cannot be excluded a priori. Therefore, we measured the variation of the intensity of the (300) reflection, $I_{300}(T)$, across the Néel temperature and compared it to that of the (330) magnetic reflection. Both temperature evolutions are shown in Fig. 3 as well as a power law fit. We note that $I_{300}(T)$ remains finite, albeit weak, above $T_N$ (see peak $E_A$ in Fig. 1). The origin of this residual intensity is the tail in the spectra of the neighboring resonating feature present at $E_B$ = 10.888 keV (which we associate to ATS scattering because of its weak temperature dependence).

The fitting of the temperature dependence of the diffracted intensity to a power law $I = (1 - T/T_N)^{\alpha_{hkl}}$ results in an estimate for the critical exponents, $\alpha_{330} = 0.72 \pm 0.06$ and $\alpha_{300} = 1.7 \pm 0.2$, respectively, for the two reflections. This finding establishes that the temperature dependence of $I_{300}(T)$ and $(I_{330}(T))^2$ have the same critical behavior below $T_N$ within the experimental accuracy, as $\alpha_{300} \sim 2 \times (\alpha_{330})$. The two estimated critical exponents for the temperature dependence of the two different reflections can be used to draw conclusions about the nature of the diffracted intensity for the (300) reflection, namely whether it is of magnetic or ATS origin. The (330) reflection is of magnetic origin as established in Ref. [16]. If both reflections were of magnetic origin, in the simplest case we would expect the critical exponent for both reflections to be the same. In addition, different critical exponents have been measured for odd and even rank harmonic satellite reflections in some incommensurate magnetic 3*d* systems [49, 50], which arise from an incommensurate magnetic structure and lattice distortions, respectively. Therefore, different critical exponents strongly favor an alternative origin of the diffracted intensity to a magnetic one.

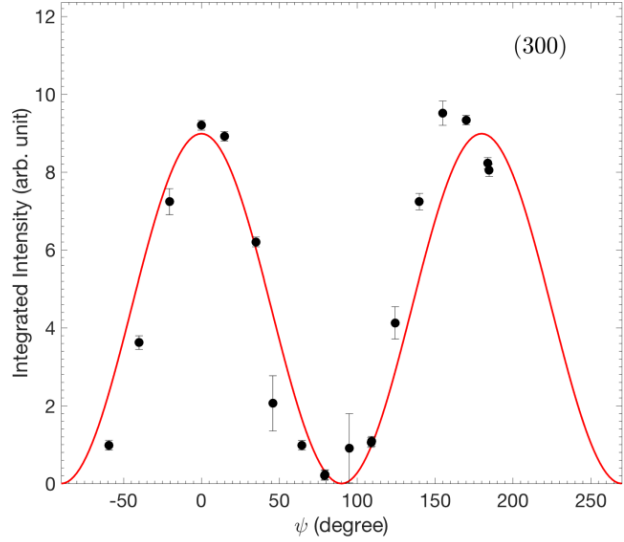

FIG. 2. *(Color online) Azimuthal angle dependence of the (300) integrated reflection intensity at 300 K acquired using the Pilatus pixel detector at the resonant energy* $E_A$ *(closed circles). The observed signal is consistent with a* $\sigma - \pi'$ *origin (continuous line).*

To completely rule out a magnetic origin in the (300) signal, we looked for intensity in the $\pi - \pi'$ polarization channel. The presence (absence) of signal in the $\pi - \pi'$ channel would be expected in the case of a magnetic (electric quadrupole) origin. Specifically, $\pi - \pi'$ measurements were carried out at several azimuths. The absence of (300) diffracted intensity was in the $\pi - \pi'$ channel, indicates the electronic (ATS) nature of the (300) reflection at the energy $E_A$. Therefore, our results suggest that, at the resonant energy $E_A$, the (300) intensity reflects an induced electronic ordering when the magnetic moments become long range ordered.

Having experimentally established the electronic origin of the (300) reflection, we can now compare the temperature evolution of the spectral features of (300) ATS reflection at energies $E_A$ and $E_B$. By comparing the absorption spectra with the resonant diffraction one (see Fig. 1), we see that the energy $E_A$ corresponds to electronic transitions to unoccupied states next to the Fermi energy, while $E_B$ corresponds to electronic transitions to excited states lying in the (unoccupied) conductive band. Such excited states are less affected by the occurrence of the phase transition, because their occupation level is not strongly affected by the temperature increase. Therefore the resonant x-ray diffraction spectrum at $E_B$ is not strongly affected by the temperature increase as seen in Fig. 1, whereas the electronic states near the Fermi level show, at least for the (300) reflection, a significant change. This change could reflect the opening of an insulating gap, which is known to occur at $T_N$ [19].

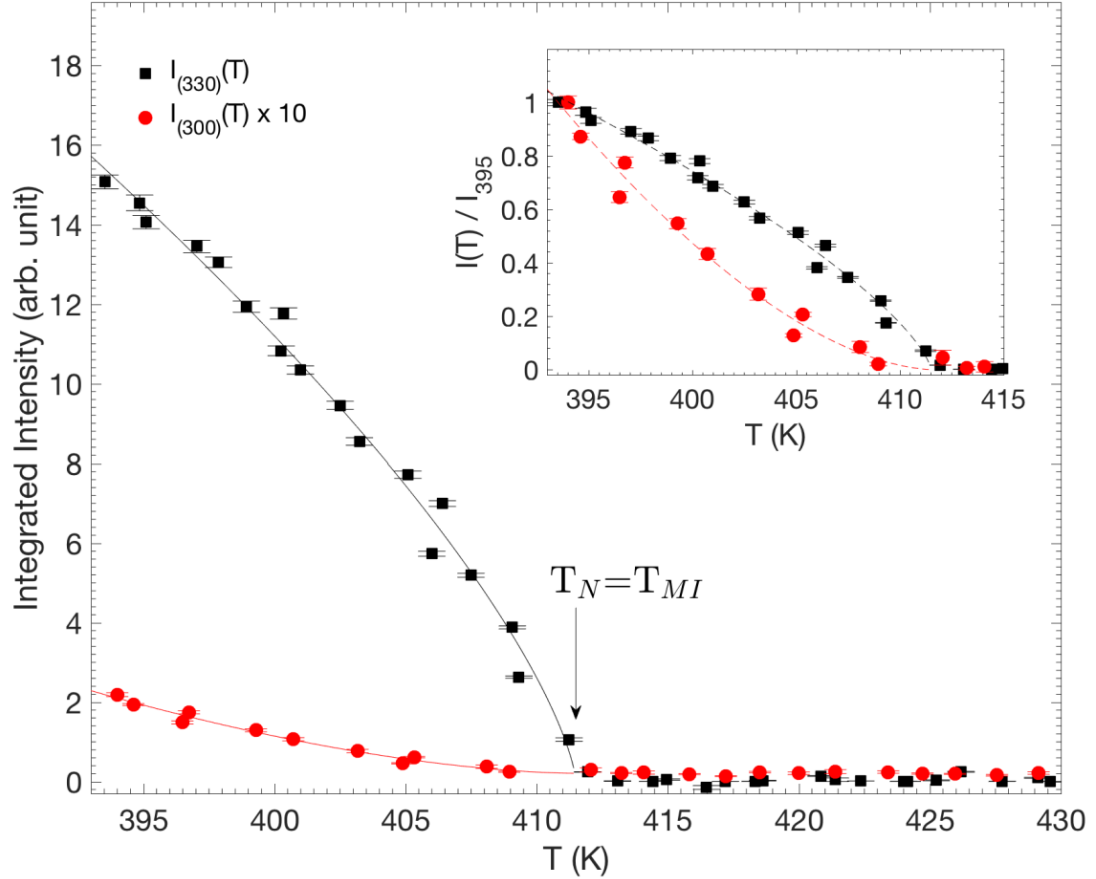

FIG. 3. *(Color online) Comparison of the temperature dependent intensity of the (300) ATS and (330) magnetic reflections at an x-ray photon energy corresponding to $E_A$= 10.878 keV. The estimated critical exponents, $\alpha_{hkl}$ for the two reflections are $\alpha_{330} = 0.72 \pm 0.08$ and $\alpha_{300} = 1.7 \pm 0.1$, respectively. In the inset, the same data in the vicinity of $T_N$ and normalized by the intensities of the respective reflections measured at T=395 K is shown, demonstrating the critical behavior of the two reflections.*

## V. FIRST-PRINCIPLES CALCULATIONS

### a. FDMNES Calculations

The original model proposed by Slater for a metal-insulator transition does not require the presence of electron correlation effects or the presence of the spin-orbit interaction. To ascertain whether the $NaOsO_3$ metal-insulator transition follows the Slater mechanism or not, we make a comparison between the experimentally observed change in the electron density from our resonant diffraction experiment with first-principle calculations in which the influence of the spin-orbit interaction and the electron correlations can be optionally taken into account. This possibility is offered by the FDMNES package [29, 51, 52] which is based on density functional theory (DFT) approach. Taking the large spatial extent of the Osmium $5d$ wave function into consideration means that the density functional theory (DFT) approach is appropriate to calculate the final states of the absorption process. We have therefore used the FDMNES package to calculate the energy dependence of the observed reflection at the Os $L_3$ and $L_2$ edges.

By comparing the FDMNES simulations with the experimental data we can ascertain if the spin-orbit interaction and electron correlations are relevant to describe the physical properties of the material. FDMNES is an ab initio code that calculates the x-ray spectroscopic response of a sample in the vicinity of an absorption edge. It requires only the crystallographic structure of the material, as well as the magnetic structure, if applicable. It is relativistic, and the spin-orbit interaction and electron correlations can be selectively taken into account by introducing the corresponding correction terms. Therefore, it is straightforward to estimate the influence of, for example, the electron correlations on the spectroscopic response of the sample.

For $NaOsO_3$, we have assumed a G-type magnetic structure with moment parallel to the $c$-axis as reported in Ref. [16] and we have disregarded the weak ferromagnetic component [15], since it is negligibly small compared to the antiferromagnetically ordered moment. With information on the crystallographic and magnetic structure, the program computes the spin-polarized electronic density of states of the specified absorbing atom surrounded by neighboring atoms within a given distance specified at the beginning of the calculation. Subsequently it calculates diffraction spectra for well-defined x-ray polarizations and Bragg wave vectors. For our simulations we have used the muffin-tin approximation, with a cluster radius around the resonant atom of 6 Å. The simulated diffraction spectra are corrected for self-absorption effects as well as the reduction in the x-ray penetration depth and the consequent reduction in the scattering volume that occurs as the photon energy is swept across the absorption edge [53].

We have performed simulations at T=300 K, 390 K and 420 K. For the simulations at T=390 K and 420 K we have used the atomic positions reported in the supplementary material of Ref. [16]. For T~300 K we have use the atomic positions in Ref. [15].

It is observed (see Fig 4) that including the effect of the spin-orbit interaction, SOI (which applies only to the final states, since the strong spin-orbit interaction for the 2$p$ core orbitals states is always accounted for by the splitting into the $L_2$ and $L_3$ manifolds), does not substantially change the absorption spectra. The green line in Fig 4 illustrates the simulated absorption spectra at the $L_3$ edge for a cluster radius of 6 Å.

We now focus our attention to the resonant diffraction spectra of the (300) and (030) reflections calculated at the $L_3$ edge. We find a good qualitative agreement with the experimental data of the (300) and (030) reflections, respectively. FDMNES also calculates a cosine dependence on the azimuthal angle for the (300) reflection, which is in agreement with the measurements.

Next, we checked how the simulations of the resonant diffraction spectra are influenced by the spin-orbit interaction, SOI. It turns out that including the spin-orbit interaction produces only subtle change in the simulations. The subtle differences being (comparing Fig 4a with Fig 4b, and Fig 4c with Fig 4d): the intensities of the diffraction spectra for 430 K metallic state is slightly lower and qualitatively in-line with what is observed experimentally. Additionally, inclusion of the presence of magnetic ordering in the simulation also does not produce a significant change in the simulated spectra. The weak influence of the presence of magnetic ordering can be related to the fact that the structural input already includes the magneto-elastic deformations originating from the presence of long range magnetic ordering.

For the calculations performed at the $L_2$ edge, the agreement with the experimental data is similar as that for the $L_3$ edge.

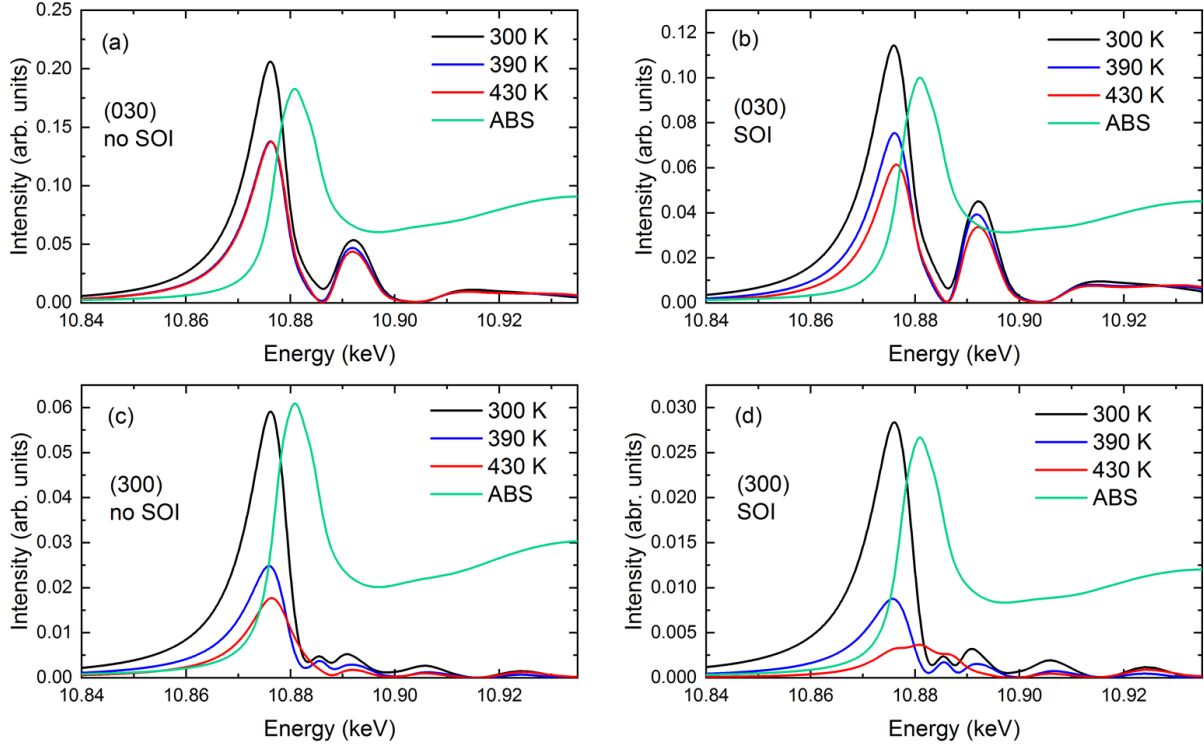

FIG. 4. *(Color online) FDMNES simulations of the temperature dependence of the intensity of the (030) reflection (top line) and of the (300) reflection (bottom line). All plots show energy scans at fixed momentum transfer at selected temperatures and x-ray absorption spectrum across the Os $L_3$ edge. All simulations were performed with cluster radius of 6 Å. Panels (a) and (b) display the simulations for the (030) reflection spectra excluding (no SOI) and including (SOI) the effect of the spin-orbit interaction (SOI). Panels (c) and (d) show simulations performed on the (300) reflection.*

At both edges, inclusion of the Hubbard correction term (which takes into account the electron correlations) also produces only subtle changes between the experimental and simulated spectra. At both $L_{2,3}$ edges, the inclusion of the spin-orbit interaction does not appreciably change the simulated spectra. We can therefore infer from our simulations that spin-orbit coupling and electron correlations play a minor role in determining the experimentally observed significant decrease in (300) intensity in the diffraction spectra across the metal-insulator transition. As a result, the (300) spectral change can be ascribed to a magneto-elastic distortion occurring in the vicinity of $T_{MI}$. However, it is not clear from the simulations if such changes are driven by the magnetic ordering directly, which would be suggestive of the Slater scenario.

One of the key experimental observations that could potentially confirm the Slater mechanism is the temperature dependence of the intensity of the (300) reflection at the $E_A$ energy, since this is proportional to the square of the intensity of the magnetic (330) peak. As this task lies beyond the current capability of FDMNES, we turn our attention to an atomic model calculation, which is described in the next Section. The advantage of such a model is that the temperature dependence of the (300) reflection naturally follows from an explicit calculation of the atomic multipole $\langle T_q^2 \rangle$ appearing in Eq (2).

**b. Atomic model calculation**

The main conclusion drawn in the previous section, namely the weakness of the electron correlation effects, motivates us to use an atomic model including multipoles, which are defined by discrete symmetries. This approach has the advantage of giving a direct insight into the physical origin of the diffracted intensity measured at Os $L$ edges.

The atomic model calculation first analyses the distortion modes driving the parent cubic structure to the observed orthorhombic one. It subsequently shows how the deformation of the cubic structure leads to the appearance of diffracted intensity, which violates the selection rules of the orthorhombic *Pnma* structure. As a next step, outlined in the Appendix B, we consider a medium coupling scheme to obtain the ground state wave-functions for the Os atom in the presence of crystal-field potential and spin-orbit coupling. Combining such results with a description of the exchange energy represented by a molecular-field, makes it possible to calculate the matrix element that gives an estimate of the quadrupole responsible for the intensity observed at the (300) reflection as well as its temperature dependence in the vicinity of the phase transition temperature. In particular, we demonstrate that an atomic model can well describe the (300) intensity including its temperature dependence, directly linking it with the appearance of the antiferromagnetic ordering.

More quantitatively, we utilize the knowledge of the magnetic space-group *Pn'ma'* to calculate unit-cell structure factors (Appendix A) and to perform a symmetry analysis of lattice distortions (Appendix B and C). In the first place, we conclude that ATS scattering is due to quadrupoles $\langle T^2_{+1} \rangle'$ at ($h$, 0, 0), $\langle T^2_{+1} \rangle''$ at ($h$, $h$, 0) and $\langle T^2_{+2} \rangle''$ at (0, $k$, 0) for $h$ and $k$ odd. Note that the quadrupole $\langle T^2_{+1} \rangle''$ is not mentioned by Calder *et al*. [16], most likely due to its negligible contribution compared to the magnetic scattering intensity from the magnetic (330) reflection.

In order to calculate an expression for the quadrupole we employ analytical techniques. It turns out that such results can be conveniently expressed taking into account the lattice deformation from the parent cubic phase, involving rotation and tilting of the $OsO_6$ octahedra. Specifically, the lattice distortions, depicted in Fig. 5, are classified as octahedral rotation and tilt angles $\theta_o$ and $\varphi_o$ (two primary order parameters), respectively, as shown in the right-hand panel of Fig. 5. As a consequence, the angles factorize in expressions for the quadrupoles (C3). The intensity of the (300) Bragg peak is successfully attributed to the first primary order parameter (octahedral rotation described by the angle $\theta_o$).

The quadrupole to be compared with experimental data is derived from,

$$\langle T^2{}_0 \rangle_T \begin{pmatrix} 1 & 0 & 0 \\ 0 & -2 & 0 \\ 0 & 0 & 1 \end{pmatrix}. \tag{B2}$$

After application of the transformation of coordinates in Eq. C1, one finds $\langle T^2{}_{+1} \rangle' \propto (xz) \propto F_{300} = \{(3/2)\sin(2\theta_o)\langle T^2{}_0 \rangle_T\}$, $\langle T^2{}_{+1} \rangle'' \propto (yz) = 0$, and $\langle T^2{}_{+2} \rangle'' \propto F_{030} \propto (xy) = 0$. The latter results mean that the origin of the (030) Bragg peak is the second primary order parameter (octahedral tilting, $\varphi_o$).

The precise structure of the spin-orbit coupling and crystal-field potential in a medium coupling scheme is derived from symmetry. Likewise, the (030) intensity is accounted for by octahedral tilting.

With such assumptions, it is possible to calculate the saturation value of the quadrupole $\langle T^2{}_0 \rangle$ and how it is related to the average magnetic moment $\langle S_c \rangle$. As outlined in the Appendix B, it can be shown that $\langle T^2{}_0 \rangle$ is proportional to $\langle S_c \rangle^2$ and that within a molecular-field calculation, $\langle S_c \rangle \propto (1 - T/T_N)^{1/2}$ as the temperature approaches $T_N$. Therefore the intensity of a Bragg peak $(h, 0, 0)$ with $h$ odd is proportional to $(1 - T/T_N)^2$, as observed experimentally.

This conclusion is also fully consistent with general symmetry arguments based on the Landau free-energy decomposition. The irreducible magnetic order parameter which drives the transition to the magnetic Pn'ma' space group is one dimensional [54] and therefore can only form either bilinear coupling term with time-odd quantity or linear-quadratic term with time-even quantity transformed by totally symmetric representation of the paramagnetic Pnma space group. No other coupling schemes are symmetry allowed. The first option imposes the critical behaviour identical with the magnetic order parameter; the second one implies a twice bigger critical exponent. The time-even ⟨$T^2_0$⟩ quadrupole does not break any symmetry and therefore is allowed to be coupled to the magnetic order parameter with the critical behaviour consistent to that observed experimentally.

We summarize briefly our findings based on our atomic model calculation in Table 1, in order to explain the temperature dependence of space group forbidden reflections. First we examine the metallic state (T>$T_N$). Above $T_N$, except for the feature $E_A$ of the (300) spectra (denoted $I_{300}^{E_A}$), all other measured forbidden reflections have an electronic ATS contribution. This contribution is due to the octahedral rotation and tilting associated with the Pnma crystallographic structure.

For the insulating state (T<$T_N$), the feature $E_B$ of the (300) spectra ($I_{300}^{E_B}$) and the (030) reflection ($I_{030}$) have a weak ATS intensity and spectral shape dependence, which we ascribe to weak temperature dependence of the secondary order parameters $\langle T_1^2 \rangle'_{\theta_o}$ and

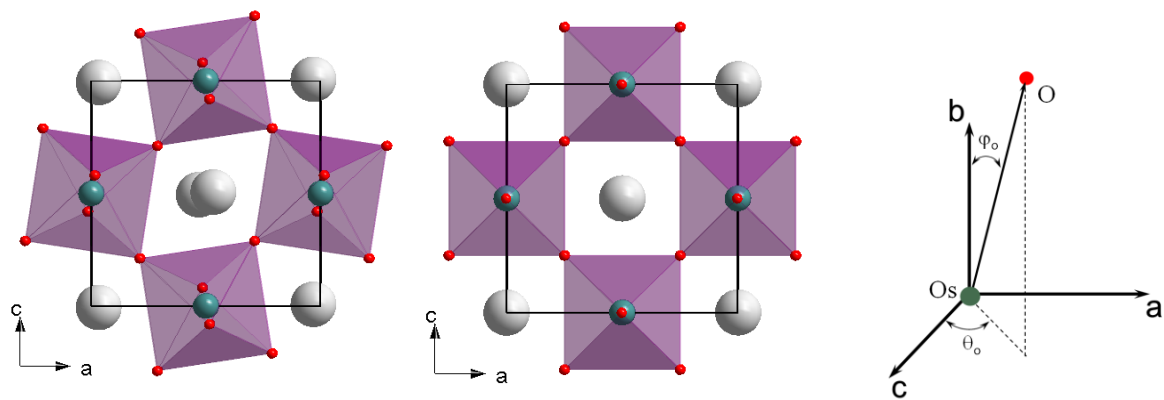

*Fig. 5. The chemical structure of NaOsO₃ (Pnma) is displayed in the left-hand panel. Positioning of almost perfect OsO₆ octahedral can be described relative to an undistorted reference structure (central panel) with the b-axis normal to the plane of the diagram. The two angles $\theta_0$ (octahedral rotation) and $\varphi_0$ (octahedral tilt) used in the text to define the b-axis in the distorted structure (left-hand panel) are illustrated in the right-hand panel. Definitions of angles that define the rotation and tilt of the O-Os-O axis of an octahedron relative to the crystal axes (a, b, c).*

$\langle T_2^2 \rangle''_{\varphi_o}$. The subscript $\theta_o$ highlights their relationship with the octahedral rotation. Conversely, the strong temperature dependence of the (330) magnetic reflection intensity $I_{330}$ below $T_N$ reflects the emergence of the antiferromagnetic ordering in the sample. Such long range ordering is described by an order parameter representing the average magnetic moment $\langle S_c \rangle$. The source of $I_{300}^{E_A}$ below $T_N$ is a magnetically induced quadrupole $\langle T^2{}_0 \rangle_T$ whose non-zero value originates from the asymmetry in the Os charge density due to the octahedral rotations. We have also shown that our model successfully describe the $I_{300}^{E_A}$ temperature dependence, which is proportional to $\langle S_c \rangle^2$.

| Reflection | Contribution T>$T_N$ | Contribution(s) T<$T_N$ | Order parameter |
|---|---|---|---|
| $(300)_{E_A}$ | none | magnetically induced ATS | $\langle T^2{}_0 \rangle_T \propto \langle S_c \rangle^2$ |
| $(300)_{E_B}$ | ATS | ATS | $\langle T_1^2 \rangle'_{\theta_o}$ |
| (030) | ATS | ATS | $\langle T_2^2 \rangle''_{\varphi_o}$ |
| (330) | ATS | Magnetism + ATS | $\langle S_c \rangle$ |

Table I. Summary of the origin of the diffracted intensity for the investigated Bragg reflections in the vicinity of the Osmium $L_{3,2}$ edges below and above the metal-insulator transition temperature $T_N$ determined from our atomic model. For each reflection, the source of the diffracted intensity above and below the Neel temperature $T_N$ is given. The last column summarizes the order parameters and, where relevant, its relation with the average ordered antiferromagnetic magnetic moment $\langle S_c \rangle$.

**VI. DISCUSSION**

The goal of this work is to clarify the nature of the metal-insulator transition in NaOsO$_3$, which has recently been proposed to be of Slater type [16]. The Slater metal-insulator transition mechanism relies on the following three conditions: (i) the lack of a significant lattice distortion and absence of crystallographic symmetry breaking, (ii) a negligible role played by electron correlation effects and (iii) the opening of the insulating gap being a result of the onset of antiferromagnetic ordering.

As for point (i), our x-ray single-crystal diffraction experiments indicate indeed that NaOsO$_3$ undergoes an isostructural metal-insulator transition across T$_N$, satisfying the Slater requirements. Condition (ii) is also satisfied as the electron correlations are weak [18, 25] and our FDMNES simulations suggest that correlations do not influence the resonant scattering intensity and can be therefore neglected. Finally, condition (iii) is compatible with our observation of the temperature dependence of the (330) and (300) diffracted intensities, which reflect the magnetic and electronic ordering, respectively. In particular, the fact that the estimated critical exponents for the temperature dependence of these two reflections are correlated, indicates that the change in the Os electron density across T$_N$ is actually magnetically induced. Therefore, our work supports the Slater scenario for the metal-insulator transition in NaOsO$_3$. The question that then arises is how exactly the antiferromagnetic ordering drives the sample into an insulating state. In order to answer this question, it is necessary to look at electronic $I_{300}^{E_A}$ diffracted intensity carefully and understand the change in its intensity across T$_N$.

From previous studies on distorted perovskite systems e.g. in Ca$_2$RuO$_4$ [55], a similar change in the intensity of a forbidden reflection was attributed to "orbital ordering", without the need for the Slater mechanism. Here, orbital ordering refers to emergence of a broken symmetry state in which localized occupied orbitals form a regular pattern, arising solely from crystallographic symmetry breaking. In systems where the change in the lattice distortions across T$_{MI}$ is too small to detect, this is referred to as an "orbital ordering" phase transition instead of an otherwise apparent structural phase transition. However, in the case of NaOsO$_3$, since there is no crystallographic symmetry breaking across T$_N$, we can exclude orbital ordering, and we need to look for an alternative physical mechanism to explain the source of the change in the electronic $I_{300}^{E_A}$ diffracted intensity across T$_N$.

The most likely alternative mechanism for the emergence of $I_{300}^{E_A}$ below T$_N$ = T$_{MI}$ in the resonant diffraction spectra is a change in the Os 5$d$ orbital population. Below T$_{MI}$ i.e. on entering the insulating phase, some of the partly occupied states become depopulated due to the opening of the insulating gap, thus allowing specific excitation channels to become available for the resonant process. In order to obtain quantitative information on the change in the Os 5*d* orbital population in the vicinity of the metal-insulator transition, it is necessary to measure the intensity of several forbidden reflections. These forbidden reflections correspond to the different components of the asymmetry of the resonant ion (Os) electron density, and thus provide a way to reconstruct its electron density across $T_{MI}$. This reconstruction process is usually quite challenging due to the large number of the intermediate states available for the core electron in the excitation process, as well as the presence of strong electron correlations in the systems. However, for the case of NaOsO$_3$, such difficulties are fortunately not so severe. This is because the electron correlations are weak in this 5*d* system and our FDMNES simulations suggest that such correlations do not influence the resonant scattering intensity and can be therefore neglected. In addition, since deformation of the OsO$_6$ octahedra is very small, a simplified analytical calculation of the quadrupoles purely allowed by symmetry can be performed. Indeed, this simplified calculation based on our atomic model explains reasonably well the temperature dependence observed for the magnetic (330) and forbidden (030) diffracted intensities in the vicinity of $T_N$. This model, also finds the electronic $I_{300}^{E_A}$ diffracted intensity is directly proportional to a single quadrupole component, $\langle T^2_0 \rangle_T$ and that its temperature dependence increases with the fourth power of the average ordered magnetic moment $\langle S_c \rangle$, as observed experimentally. Therefore, using our atomic model, we were able to reconstruct the change in the electron density occurring at the metal-insulator transition.

In order to visualize this change in electron density of the Os ion across $T_{MI}$, we plot the isosurfaces of the 5*d* electron density of Os ion above and below the $T_{MI}$ (Fig 6c and 6d, respectively). Since, the Os$^{5+}$ 5*d* shell is unfilled, its charge distribution deviates significantly from spherical symmetry in both the metallic and insulating states. In order to plot the isosurfaces of 5*d* electron density, we first need to know the number of electrons per orbital. To begin with, in a localized electron picture of the charge distribution for Os$^{5+}$, one would assume that there is one valence electron in each of the *t$_{2g}$* orbitals as dictated by Hund's rules. However, for the itinerant 5*d* system, one would expect substantial *p-d* hybridization, and a more realistic picture is given in Ref. [18], which is based on Local Spin Density Approximation (LSDA) *ab initio* calculations. These calculations performed for the insulating state in NaOsO$_3$ suggest an occupation of 0.7 electrons per orbital for majority *t$_{2g}$* orbitals and an occupation of about 0.3 electron per orbital for the majority *e$_g$* and all the minority *d* orbitals, which are formally unoccupied. We used these values as a starting point to plot an isosurface of the Os charge density in the insulating state (see Fig. 6c). In the metallic state, according to our experimental observation combined with our symmetry analysis, the $e_g$ states become thermally populated leading to a change in the charge density towards a slightly more spherical shape (see Fig. 6b) with an occupation of about 0.5 electron per orbital for all the $e_g$ orbitals, as extracted from the symmetry analysis performed in Section Vb.

By comparing Fig. 6b and Fig. 6c, it becomes clear that the Os electron density undergoes only a minor change in shape at the metal-insulator transition. This result is not surprising as the change in the lattice constants and the Osmium-Oxygen bonding angle are of the order of a fraction of a percent [16] across $T_{MI}$. Nevertheless, such small changes can substantially affect the resonant x-ray diffraction response of the sample as has been seen from our experiments in combination with a detailed symmetry analysis (Section Vb and the Appendix A). Therefore resonant x-ray diffraction is an ideally suited technique to detect even the slightest changes occurring in the electron density of the resonant ion across $T_{MI}$.

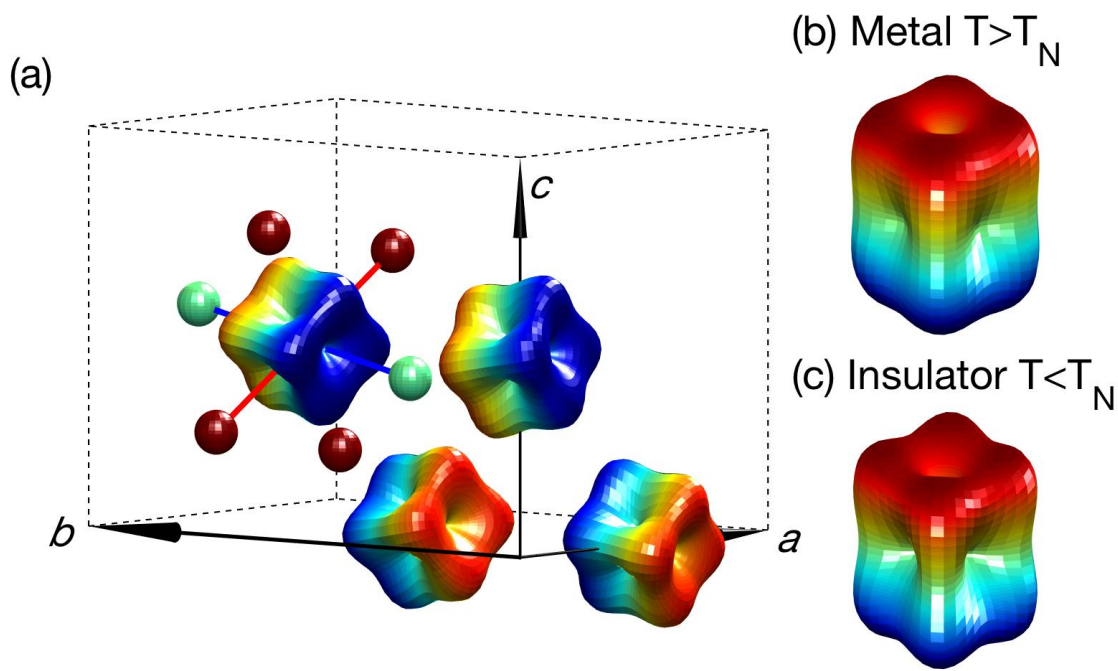

FIG. 6. (Color online) (a) Isosurface of the 5d electron density of Os ions in the insulating state of NaOsO$_3$. The isosurface colors highlight the symmetry relation between the Os ions in the unit cell, i.e. Os ions sitting at (0 0 1/2) and at (1/2 0 0) positions are related by a rotation about the c-axis of 180°. For clarity only selected Os atom and oxygen (represented as green and red spheres for Wyckoff positions 4c and 8d respectively) are shown. (b) and (c) are the reconstructed 5d electron density distribution for the metallic and the insulating phases, respectively.

## VII. CONCLUSIONS

From our single crystal non-resonant and resonant x-ray diffraction experiments, we establish that $NaOsO_3$ fulfills the prerequisites of a Slater metal-insulator transition as follows: first, no crystallographic symmetric breaking occurs across $T_{MI}$. Second, the presence of weak electron correlations in the 5$d$ $NaOsO_3$ system has been confirmed by the *ab initio* calculations performed with the FDMNES software. Third, the antiferromagnetically driven change of Os electron density across $T_N = T_{MI}$ results in the opening of the insulating gap. Specifically, it is the appearance of the electronic (300) diffracted intensity below $T_N$ that reflects the change in the Os 5$d$ orbital population across $T_{MI}$, which has been well explained by our theoretical atomic model.

To conclude, our results favor the Slater mechanism as the origin of the metal-insulator transition in $NaOsO_3$.

## ACKNOWLEDGEMENTS

We are grateful to K. S. Knight, A. Gibbs, G. van der Laan, M. Glazer, M. Arai and S. di Matteo for fruitful discussions and to D. Chernychov for help during the experiment at the Swiss-Norwegian beamline at the ESRF. We also wish to thank L. Rettig and W. Y. Windsor helped in preliminary sample characterization work that was carried out at the MS beamline of the Swiss Light Source, Paul Scherrer Institute, Villigen, Switzerland. We thank Diamond Light Source for access to beamline I-16 that contributed to the results presented here. The research leading to these results has received funding from the Swiss National Science Foundation. Work in Japan was supported by JSPS KAKENHI Grant Numbers JP15K14133 and JP16H04501. The raw data and the FDMNES simulation files that support this study are available via the Zenodo repository at https://doi.org/10.5281/zenodo.1217788

## VIII. APPENDIX A

X-ray and neutron Bragg diffraction experiments have established the chemical structure *Pnma* (#62) for NaOsO$_3$, which is often the case for perovskite oxides, with Os ions using sites 4$a$ that possess inversion symmetry, $\bar{1}$, and no more. The superstructure is due to the cooperative rotation and tilting of the OsO$_6$ octahedra as in the GdFeO$_3$-type perovskite. In this appendix we outline the derivation of exact unit-cell structure factors for ($h$00), (0$k$0) and ($hh$0) Bragg peaks.

An appropriate electronic structure factor is,

$$\Psi^K{}_Q = \sum_\mathbf{d} \exp(i\mathbf{d} \cdot \boldsymbol{\tau}) \langle T^K{}_Q \rangle_\mathbf{d}, \tag{A1}$$

where the Bragg wavevector $\boldsymbol{\tau} = (h, k, l)$, and sites labelled $\mathbf{d}$ in a cell are occupied by Os ions. Universal expressions for $F_{\mu'\nu}$ ($E1 - E1$) are written in terms of two quantities $A^K{}_Q = \{(\Psi^K{}_Q + \Psi^K{}_{-Q})/2\}$ and $B^K{}_Q = \{(\Psi^K{}_Q - \Psi^K{}_{-Q})/2\}$.

The metal-insulator transition proceeds without change to the crystal symmetry, and the G-type motif of Os magnetic dipoles possesses a commensurate propagation vector $\mathbf{k}$ = (0, 0, 0). Data in Fig. 3 of reference [15] are evidence of weak ferromagnetism. Os ions use sites 4a with site symmetry $\bar{1}$ in Pn'ma' and they are,

$$(0, 0, 0)_1 \uparrow: (1/2, 1/2, 1/2)_2 \uparrow: (0, 1/2, 0)_3 \downarrow: (1/2, 0, 1/2)_4 \downarrow,$$

where arrows indicate relative orientations of dipole moments along the c-axis inferred from diffraction data [16]. Sites (1) & (3) and (2) & (4) are related by an anti-translation (0, 1/2, 0)', while (1) & (2) and (3) & (4) are related by a simple body-centre translation (1/2, 1/2, 1/2).

With this information one finds,

$$\Psi^K{}_Q = [1 + \sigma_\theta (-1)^{h+l}(-1)^Q]$$
$$\times [\langle T^K{}_Q \rangle + (-1)^{h+k+l}\sigma_\theta (-1)^K \langle T^K{}_{-Q} \rangle], \tag{A2}$$

where $h$, $k$ & $l$ are integer Miller indices and $\sigma_\theta$ is the time signature. Bulk properties are prescribed by $\Psi^K{}_Q$ evaluated for $h = k = l = 0$. By taking $\sigma_\theta = -1$ a ferromagnetic motif of dipoles (K = 1) parallel to the b-axis is allowed.

The result (A2) applies to the calculation of a unit-cell structure factor for magnetic neutron diffraction on using $\sigma_\theta = -1$. Bragg peaks are attributed to magnetic dipoles. Reflections have been indexed by extinction rules $k$ odd and $h + l$ odd, and they refer to a magnetic moment $= (0, 0, \mu_c)$ [16].

Henceforth, we focus attention on the interpretation of x-ray Bragg diffraction with signals enhanced from tuning the primary energy to the energy of a parity-even absorption event. In this instance, $\sigma_\theta = (-1)^K$ and $\Psi^K_Q = (-1)^{h+k+l}\, \Psi^K_{-Q}$, while the first bracket in (A2) demands that $(K + Q + h + l)$ is even. Hence, a motif of antiferromagnetic dipoles ($K = 1$) alternating along the c-axis ($Q = 0$) can be observed with $k$ odd and $h + l$ odd, in accord with neutron diffraction data [16]. Taking Q odd yields the rule $h + l$ even for magnetic multipoles with K odd. In this case,

$$\Psi^K_Q = 2\,[\langle T^K_Q \rangle + (-1)^k \langle T^K_{-Q} \rangle] = 2\,[\langle T^K_Q \rangle - (-1)^k \langle T^K_Q \rangle^*]. \tag{A3}$$

Applied to dipoles ($K = 1$, $Q = \pm 1$), this result says that antiferromagnetic moments along the a-axis are visible at $k$ odd ($\Psi^1_a = -4\sqrt{2}\, \langle T^1_{+1}\rangle'$) while ferromagnetic moments along the b-axis are visible at $k$ even ($\Psi^1_b = -4\sqrt{2}\, \langle T^1_{+1}\rangle''$). Note that $\Psi^1_a$ and $\Psi^1_b$ are purely real. Turning attention to other components, one finds $(\Psi^K_Q + \Psi^K_{-Q}) = 0$ for $(h + k + l)$ odd.

The last result means $A^K_Q = 0$ for $(h, 0, 0)$ with $h$ odd. In consequence, corresponding unit-cell structure factors for unrotated polarizations $F_{\sigma\sigma'} = F_{\pi\pi'} = 0$. The allowed $B^K_Q$ possess $(K + Q)$ odd, with $B^K_0 = 0$ by definition. We find $B^2_1 = (4\langle T^2_{+1}\rangle')$ and,

$$F_{\sigma\pi'}\,(E1\text{-}E1) = \cos(\theta)\,\cos(\psi)\,\langle T^2_{+1}\rangle'. \tag{A4}$$

In this $(h, 0, 0)$ structure factor, and all subsequent structure factors, we drop a factor 4 that comes from the number of Os ions in a unit cell. The origin of the azimuthal angle scan ($\psi = 0$) finds the c-axis and magnetic moments normal to the plane of scattering.

Structure factors for $(h, 0, 0)$ that we have considered are directly proportional to $\Psi^K_Q$ because the Bragg wavevector is parallel to the x-axis as depicted in Fig. 1 of Ref. [36] that defines states of polarization in the primary and secondary beams. Hence, $B^2_1 = -\Psi^2_1$ in the derivation of (4) with a minus sign due to rotation by 180° about the c-axis. For $(0, k, 0)$ we must allow for a rotation of $\Psi^K_Q$ by 90° about the c-axis,

$$A^K_Q = i \sin(\pi Q/2) \, \Psi^K_Q \text{ and } B^K_Q = \cos(\pi Q/2) \, \Psi^K_Q, \tag{A5}$$

for $k$ odd. Non-vanishing $\Psi^K_Q$ possess $(K + Q)$ even, and E1-E1 structure factors are determined by $A^1_1 = -(i\sqrt{2}\langle T^1_a \rangle)$ and $B^2_2 = -(2i \langle T^2_{+2} \rangle'')$. For $(0, k, 0)$ with $k$ odd one finds $F_{\sigma'\sigma} = 0$,

$$F_{\sigma\pi'} (E1\text{-}E1) = -i \cos(\theta) [(1/\sqrt{2}) \cos(\psi) \langle T^1_a \rangle + i \sin(\psi) \langle T^2_{+2} \rangle''], \tag{A6}$$

$$F_{\pi\pi'} (E1\text{-}E1) = -(i/\sqrt{2}) \sin(2\theta) \sin(\psi) \langle T^1_a \rangle.$$

Time-even, $\langle T^2_{+2} \rangle''$, and time-odd, $\langle T^1_a \rangle$, multipoles in the rotated channel of polarization differ by a 90° phase. However, the magnetic moment is observed to be parallel to the c-axis leading us to expect $\langle T^1_a \rangle \approx 0$.

The third, and last, set of Bragg peaks we consider are indexed by $(h, h, 0)$ with $h$ odd [16]. From (A2), $\Psi^K_Q = \Psi^K_{-Q}$ and the rank and projection of non-vanishing $\Psi^K_Q$ are constrained by $(K + Q)$ odd. One needs,

$$A^K_Q = \cos(\varphi Q) \, \Psi^K_Q \text{ and } B^K_Q = i \sin(\varphi Q) \, \Psi^K_Q. \tag{A7}$$

to evaluate structure factors, which we choose to report as functions of $A^K_Q$ and $B^K_Q$ to ease complexity of the notation. Using unit cell lengths $a \approx 5.392$ Å, $b \approx 7.608$ Å [16] we find $\varphi \approx 144.67°$. Of the three contributions to E1-E1 unit-cell structure factors, $A^1_0$ and $B^2_1$ are purely real and $A^2_1$ is purely imaginary, with $A^2_1 \propto B^2_1 \propto \langle T^2_{+1} \rangle''$. We find,

$$F_{\sigma\sigma'} (E1\text{-}E1) = -i \sin(2\psi) A^2_1,$$

$$F_{\sigma\pi'} (E1\text{-}E1) = -(i/\sqrt{2}) \cos(\theta) \sin(\psi) A^1_0 - i \sin(\theta) \cos(2\psi) A^2_1 - \cos(\theta) \cos(\psi) B^2_1,$$

$$F_{\pi\pi'} (E1\text{-}E1) = (i/\sqrt{2}) \sin(2\theta) \cos(\psi) A^1_0 - i \sin^2(\theta) \sin(2\psi) A^2_1. \tag{A8}$$

The dipole $A^1_0 \propto \langle T^1_0 \rangle$ does not contribute to diffraction in the rotated channel of scattering at the origin of the azimuthal-angle scan. When the c-axis is in the plane of scattering ($\psi = 90°$) the structure factor $F_{\pi'\sigma}(E1\text{-}E1)$ is a combination of $\langle T^1_0 \rangle$ and $\langle T^2_{+1} \rangle''$, and $F_{\sigma\sigma'} (E1\text{-}E1) = 0$. Calder *et al*. [16] report the result $F_{\sigma\sigma'} (E1\text{-}E1) = 0$ and it is likely they used $\psi = 90°$. In which case, data for intensity in the rotated channel are caused by the magnetic dipole and, also, a non-magnetic quadrupole, $\langle T^2_{+1} \rangle''$, that was previously omitted from interpretations of data. Note that $(h, 0, 0)$ Bragg peaks are created by the real part of the same quadrupole, i.e., $\langle T^2_{+1} \rangle'$.

**APPENDIX B**

In this Appendix we derive a quantitative expression for the quadrupole $\langle T^2_0\rangle_T$, which is at the origin of the diffracted signal observed at the (300) Bragg peak. In order to do so, we first note that besides the large tilting and rotation of the OsO$_6$ octahedra, causing the departure from cubic symmetry, the octahedral preserve their shape and are almost perfect. The variation of the six bond distances between O and Os is smaller than 0.4% of the largest distance, and the O–Os–O angles are 90º, to a very good approximation [15]. The large orthorhombic distortion leaves a hybridization gap of ≈ 1.6 eV between $t_{2g}$ and $e_g$ antibonding manifolds [26], in addition to reducing the $t_{2g}$ bandwidth [15]. Hence, local electronic structure can safely be assigned to equally occupied $t_{2g}$ orbitals retaining local cubic symmetry in the OsO$_6$ octahedron.

The axes of an octahedron depart substantially from the crystal axes, and the corresponding distortion, in the form of tilt and rotation of the O-Os-O axis have values $\theta_o \approx \varphi_o \approx$ 10° [15]. Taking $\theta_o$ = 0° and $\varphi_o$ = 0° returns to the undistorted structure with orthonormal axes *(x$_o$, y$_o$, z$_o$)* transforming to *(a, b, c)* with,

$$x_o = (cos\theta_o,\ 0,\ -sin\theta_o),$$

$$y_o = (sin\varphi_o\ sin\theta_o,\ cos\varphi_o,\ sin\varphi_o\ cos\theta_o), \tag{B1}$$

$$z_o = (cos\varphi_o\ sin\theta_o,\ -sin\varphi_o,\ cos\varphi_o\ cos\theta_o).$$

Our analysis of quadrupoles in *Pn'ma'* that contribute to ATS scattering shows that the octahedral rotation accounts for the (300) intensity. This is specifically due to the octahedral rotation in tetragonal *P4/mbm* with Os ions at sites *2a* that possess site symmetry *4/m* on the *b*-axis (Appendix C). The quadrupole to be compared with experimental data is derived from,

$$\langle T^2_0\rangle_T \begin{pmatrix} 1 & 0 & 0 \\ 0 & -2 & 0 \\ 0 & 0 & 1 \end{pmatrix}. \tag{B2}$$

This form of the quadrupole is consistent with site symmetry *4/m* on the *y*-axis, with elements $(xx) = (zz)$ while the element $(yy)$ obeys $\{(xx) + (yy) + (zz)\} = 0$. After application of the transformation of coordinates in Eq. C1, one finds $\langle T^2_{+1}\rangle' \propto (xz) \propto F_{300} = \{(3/2)\ sin\ (2\theta_o)\ \langle T^2_0\rangle_T\}$, $\langle T^2_{+1}\rangle'' \propto (yz) = 0,$ and $\langle T^2_{+2}\rangle'' \propto F_{030} \propto (xy) = 0$. The latter results means that the origin of the (030) Bragg peak is the second primary order parameter (octahedral tilting, $\varphi_o$ ). In this case, Eq. (B2) is replaced by a quadrupole (*C*2) that explicitly displays the lower site symmetry *2/m* on the *x*-axis contained in orthorhombic *Imma.* Lower symmetry allows off-diagonal elements in the quadrupole and non-vanishing values for elements $(xz)$, $(yz)$ and $(xy)$. However, the value of $(xz)$ from octahedral tilting is not a natural source of the temperature dependence of the (300) Bragg peak as we successfully argue in the remainder of this subsection.

In our model, it is assumed that an admixture of excited terms due to the spin-orbit interaction can be neglected, so that we are concerned only with the ground term $^4$F of pentavalent Os. Furthermore, the spin-orbit interaction is taken to be appreciably stronger than the Jahn-Teller coupling and quenches the latter, to a good approximation.

In octahedral symmetry, the ground-state of $d^3$ is a half-filled $t_{2g}$ shell with a high-spin configuration $S = 3/2$ and L = 3 ($^4$F). Some previous studies of the $d^3$ configuration in cubic symmetry were motivated by $Cr^{3+}$ in $Al_2O_3$ (pink ruby) [56]. This configuration is robust against the strength of the crystal-field energy relative to the spin-orbit coupling, which determines Hund's first rule. If the spin-orbit interaction is the dominant force in determining electronic states of an Os ion, then the total angular momentum *J* is a good quantum number. Atomic states are $|J, \pm 1/2\rangle$, $|J, \pm 3/2\rangle$ with $J = 3/2$ for $^4F_{3/2}$ ($d^3$). This coupling scheme gained favor in a simulation of electronic structure [57], but our measurements on $NaOsO_3$ rule against it for this material. Calculations of the quadrupole $\langle T^2_0 \rangle$ using $J = 3/2$ in $E1 - E1$ diffraction amplitudes for $L_3$ and $L_2$ absorption edges differ by a factor 10 ($\langle T^2_0 \rangle_{L_2}/\langle T^2_0 \rangle_{L_3} = -10$), resulting in a factor 100 difference in intensity. In contrast, our measured intensities are roughly equal at the two edges, and even if we consider the correction of intensity by absorption effects, this would lead only to a lower bound of 1/5 in the ratio. Thus, we consider a medium coupling scheme, which takes into account the crystal-field potential and spin-orbit coupling, and find $L_3$ and $L_2$ intensities that match our measurements.

The ground state of the crystal-field potential is an orbital singlet $|\Gamma_2\rangle = [|+2\rangle - |-2\rangle]/\sqrt{2}$ with states $|M\rangle = |L = 3, M\rangle$. The orbital angular momentum in $|\Gamma_2\rangle$ is fully quenched, but it is made non-vanishing by a relatively large spin-orbit interaction ($\lambda \mathbf{S} \cdot \mathbf{L}$) with positive $\lambda$. Likewise, quadrupoles that create Templeton-Templeton scattering at space-group forbidden reflections are proportional to $\lambda$.

In this coupling scheme, the ground state of $^4$F belongs to the quadruplet $\Gamma_8$ representation of the double cubic group (the notation U' is also used for this representation [58]). The representation of the group spanned by the components of a dipole is $\Gamma_4$. Non-vanishing matrix elements of a dipole in $\Gamma_8$ are allowed if the direct product $\Gamma_8 \times \Gamma_4$ contains $\Gamma_8$, and $\Gamma_8 \times \Gamma_4 = (\Gamma_6 + \Gamma_7 + 2\Gamma_8)$. The fact that $\Gamma_8 \times \Gamma_4$ actually contains $\Gamma_8$ twice indicates that allowed matrix elements are defined by two factors. (This result is usefully contrasted with the direct product $\Gamma_5 \times \Gamma_4$, say, that contains $\Gamma_5$ once. As a consequence, a dipole in $\Gamma_5$ can be mapped to a fictitious angular momentum using one unique proportionality factor.). Moreover, the $\Gamma_4$ component of the direct product $\Gamma_8 \times \Gamma_8$ is symmetric, and thus the dipole must be time-odd for an odd number of electrons [58-63]. The representations of the group spanned by the components of a quadrupole are $\Gamma_3 + \Gamma_5$. The direct product $\Gamma_8 \times \Gamma_3$ contains $\Gamma_8$ once, while $\Gamma_8 \times \Gamma_5$ contains $\Gamma_8$ twice. However, the quadrupole $T^2_0$ that is required is unique in $\Gamma_3$ leaving one proportionality factor to be determined by an explicit calculation.

We denote by $|\Sigma\rangle$, with $\Sigma = \pm 1/2, \pm 3/2$, the four partners of the quadruplet $\Gamma_8$, which can be viewed as replacements for $|J, \pm m\rangle$ used in the extreme case of a dominant spin-orbit interaction. Keeping in mind the aim to find an expression for the exchange energy acting on an Os ion in mind, we choose

$\langle \Sigma = 1/2|S_z|\Sigma = 1/2\rangle = v$, and $\langle \Sigma = 3/2|S_z|\Sigma = 3/2\rangle = u$,

where $u$ and $v$ are real positive numbers with $u > v$ (actual values of the parameters $u$ and $v$ are inferred from measurements. Formally, however, they can be related to crystal-field energies and eigenvalues calculated by Lea, Leask and Wolf [64]. With this line of reasoning, $u$ and $v$ are functions of a parameter $x$ used by the authors to quantify energies and eigenvalues). A singular property of the $\Gamma_8$ manifold is that a matrix element of a transverse component of a dipole taken between $\Sigma = 3/2$ and $\Sigma = -3/2$ can be different from zero. Applied to the dipole operator $(\boldsymbol{L} + 2\boldsymbol{S})$ this property results in Zeeman states not being equally spaced. The quadrupole $T^2_0$ is represented by the operator $\{3(\Lambda_z)^2 - 15/4\}$ that transforms like $\Gamma_3$, where $(\Lambda_z|\Sigma\rangle) = \Sigma |\Sigma\rangle$ and the constant $15/4 = (3/2)(5/2)$ is the one expected for an operator space with maximum projections $\Sigma = \pm 3/2$. The proportionality factor in $T^2_0 \propto \{3(\Lambda_z)^2 - 15/4\}$ is to be determined, in the same way that a reduced matrix-element in the Wigner-Eckart theorem must be determined by an explicit calculation.

Identities $\langle \Sigma = -1/2|S_z|\Sigma = -1/2\rangle = -v$ and $\langle \Sigma = -3/2|S_z|\Sigma = -3/2\rangle = -u$ flow from the equivalence of a diad axis of rotation symmetry on the y-axis of the cubic group and time-

reversal symmetry. A tetrad axis of rotation symmetry on the z-axis forbids off-diagonal matrix elements of $S_z$. Likewise, $\langle \Sigma | T^2_0 | \Sigma \rangle$ is independent of the sign of $\Sigma$ while,

$$\langle \Sigma = 1/2 | T^2_0 | \Sigma = 1/2 \rangle = - \langle \Sigma = \pm 3/2 | T^2_0 | \Sigma = \pm 3/2 \rangle.$$

Middey et al. [20] conclude from their simulations that $t_{2g}$ orbitals are split by a super-exchange due primarily to nearest-neighbour Os ions. Our exchange energy is represented by a molecular-field value $(-J \cdot \langle S_c \rangle \cdot S_c)$ with $J$ the super-exchange parameter. As a consequence, the thermal average of the quadrupole in $\Gamma_8$ symmetry is,

$$\langle T^2_0 \rangle_T = - \langle T^2_0 \rangle \, (2/Z) \, \{\cosh(Au) - \cosh(Av)\}, \tag{B3}$$

where T is the temperature, $A = \{J \langle S_c \rangle/(T\sqrt{2})\}$ and

$$Z = 2 \, [\cosh(Au) + \cosh(Av)], \tag{B4}$$

is the partition function. The factor $(1/\sqrt{2})$ in $A$ arises because the molecular field is aligned with the crystal $c$-axis, whereas states $|\Sigma\rangle$ are defined in axes of the reference structure depicted in the center panel of Fig. 5. Evidently, $-\langle T^2_0 \rangle$ is the saturation value of $\langle T^2_0 \rangle_T$. For small $A$,

$$\langle T^2_0 \rangle_T \approx - \langle T^2_0 \rangle \, \{(u^2 - v^2)/(u^2 + v^2)^2\} \, \langle S_c \rangle^2. \tag{B5}$$

The spin moment is derived from,

$$\langle S_c \rangle = (\sqrt{2}/Z) \, \{u \sinh(Au) + v \sinh(Av)\}, \tag{B6}$$

which leads to,

$$\langle S_c \rangle \propto (1 - T/T_N)^{1/2},$$

as the temperature approaches $T_N = \{(u^2 + v^2)) J/(2\sqrt{2})\}$. The intensity of a Bragg peak $(h, 0, 0)$ with $h$ odd is proportional to $(1 - T/T_N)^2$, according to the molecular-field calculation.

A calculation of the proportionality factor $\langle T^2_0 \rangle = \langle \Sigma = 1/2 | T^2_0 | \Sigma = 1/2 \rangle$ is aesthetically pleasing, because it demonstrates that $\langle T^2_0 \rangle$ arises from spin-orbit interaction in the medium coupling scheme. To this end, we derive approximations to $|\Sigma\rangle$ using perturbation theory. A $d^3$ ground state $|\Gamma_2; \sigma\rangle$ is a product $(|S, \sigma\rangle \, |\Gamma_2\rangle)$ using an orbital singlet $|\Gamma_2\rangle$. Specifically, orbital

angular momentum acquires a value by spin-orbit mixing of $|\Gamma_2; \sigma\rangle$ and $|\Gamma_5, \alpha\rangle$ that are separated by an energy $\Delta_o$, and a result $\langle\Gamma_2; \sigma| L_z |\Gamma_2; \sigma\rangle \approx - (8\lambda/\Delta_o)\sigma$ is derived from first-order perturbation theory (an apparently similar calculation of $\langle L_z\rangle$ for pentavalent Os in pyrochlore-type $Cd_2Os_2O_7$ omits the contribution linear in $\lambda$ that we report, and use of an erroneous relation $\langle L_z\rangle \propto \lambda^2$ understandably yields a misleading interpretation of dichroic signals [65]).

One finds $\langle\Gamma_2; \sigma|T^2_0|\Gamma_2; \sigma\rangle = 0$ for all spin projections $\sigma$, as already mentioned. A spin-orbit interaction creates admixtures of $|\Gamma_2; \sigma\rangle$ and $|\Gamma_5; \sigma'\rangle$, where $|\Gamma_5\rangle$ is triply degenerate in perfect cubic symmetry (orbital states $|\Gamma_2\rangle$ and $|\Gamma_5\rangle$ are separated by an energy 10 Dq). Addition of a tetragonal distortion reduces the degeneracy of $|\Gamma_5\rangle$ to a singlet state $|\Gamma_5, \alpha\rangle = \{|+2\rangle + |-2\rangle\}/\sqrt{2}$, and a pair of Kramers' degenerate states, one component of which is $|\Gamma_5, \beta\rangle = [\sqrt{(5)} |+1\rangle - \sqrt{(3)} |-3\rangle]/\sqrt{8}$. All three $\Gamma_5$-states mix with $|\Gamma_2; 1/2\rangle$. An off-diagonal matrix element of $T^2_0$ between $|\Gamma_2; 1/2\rangle$ and $|\Gamma_2; -1/2\rangle$ is zero, because the two states are related by time-reversal and this quadrupole is time-even. Likewise, diagonal matrix elements of $T^2_0$ are identical. After a lengthy calculation, we obtain,

$$\langle T^2_0\rangle \approx - 14\sqrt{(2/3)} (\lambda/\Delta) \langle\Gamma_5, \beta; 3/2| T^2_0 |\Gamma_2; 1/2\rangle - 2 (\lambda/\Delta_o) \langle\Gamma_5, \alpha; 1/2| T^2_0 |\Gamma_2; 1/2\rangle$$

$$= - (\pm) (\lambda/45) \sqrt{(2/3)} [(7/\Delta) + (1/\Delta_o)], \qquad (B7)$$

which is correct to leading order in $\lambda$. An energy difference $(\Delta - \Delta_o)$ is created by a tetragonal addition to the crystal-field potential. The negative sign in $\langle T^2_0\rangle$ applies at the $L_3$ absorption edge and the plus sign applies at $L_2$.

Several resources are used in the derivation of (B7) in addition to applications of perturbation theory for $|\Sigma\rangle$. First, the reduced matrix element of a quadrupole operator (Eq. (73) in Ref. [35]). Therein unit tensors $W^{(1, 1)}$ and $W^{(1, 3)}$ calculated with fractional parentage coefficients for hole states in the d-configuration [66], specifically equation (3.8) and Table 1. Unit tensors are reduced matrix-elements of specific operators, e.g., $\boldsymbol{S} \bullet \boldsymbol{L}$ is an operator equivalent for $W^{(1, 1)}$. Matrix elements in Russell-Saunders coupling scheme are best calculated with an identity (D.1) in reference [35].

The saturation value of the magnetic moment $\mu_c = \langle L_c + 2S_c\rangle \approx [2 - (8\lambda/\Delta_o)](u/\sqrt{2})$ and the orbital moment opposes the spin moment in agreement with Hund's third rule. Hence, an increase in the coupling parameter $\lambda$ diminishes $\mu_c$, a behaviour which accords with a comprehensive simulation of the influence of the spin-orbit interaction on electronic structure [57]. Using $(8\lambda/\Delta_o) = 0.2$, which is a typical value for orbital angular momentum induced by the crystal-field potential, the observed saturation moment $\mu_c \approx 1.0$ implies $u \approx 0.79$ [16] (published estimates of $\langle L_c \rangle / \langle 2S_c \rangle$ include $\approx -0.12$ [17] and $\approx -0.09$ [18]). A large covalency, which is anticipated in simulations of electronic structure showing strongly hybridized Os $5d$ and O $2p$ states [17, 18], with antiferromagnetic order causes problems for a quantitative determination of magnetic moments by neutron diffraction. Hubbard and Marshall discuss various consequences of covalent bonding depending on local crystal symmetry [67]. These include an apparent loss of the magnetic moment through full cancellation of unpaired spins at ligands, due to their symmetric placement with respect to antiparallel moments on metal ions as in magnetic $NaOsO_3$ or, conversely, magnetic contamination of nominally nuclear Bragg peaks from less than full cancellation of unpaired spins at ligand sites. A substantial reduction of $u$ from its nominal value can stem from strong p-d hybridization, evidence for which is found in results recorded by Jung et al. [18] from their simulation of insulating $NaOsO_3$, with a uniform 30% reduction in the occupation of $t_{2g}$ orbitals, and a similar magnitude for occupations of orbitals that are formally unoccupied is found. These findings support the use of $u$ and $v$ as empirical quantities. Another small reduction in the magnetic moment comes from the rotation and tilt of the octahedron, which is exemplified by the difference between the two structures depicted in Figure 5; the reduction factor is ($cos\theta_o \, cos\varphi_o$) which is likely to be a 3% effect.

**APPENDIX C**

We have successfully argued that diffraction at the space-group forbidden reflection (300) can be attributed to an octahedral rotation using *P4/mbm*. Likewise, diffraction at (030) is accounted for by octahedral tilting using *Imma*. Here, we present the supporting case for this assertion and provide additional details of the mode analysis for the distortions, namely rotation and tilting.

The orthorhombic *Pnma* structure is illustrated in Fig. 5; the distorted structure that fits $NaOsO_3$ is depicted in the left-hand panel, and it is derived from a reference structure in the center panel by rotation and tilting of almost perfect $OsO_6$ octahedra. Angles $\theta_o$ and $\varphi_o$ that quantify the rotation and tilting are shown in the right-hand panel and defined specifically in equations (B2). These distortions represent two distinct order parameters associated with $M^{3+}$ (rotation) and R$^{4+}$ (tilting) irreducible representations of the parent cubic $Pm\bar{3}m$ (#221) space group. It is useful to decompose the orthorhombic structure $Pnma$ in terms of these primary order parameters and consider them separately. This approach reveals the distortions relevant to electronic properties discussed in the manuscript. The M$^{3+}$ ($\eta_M, 0, 0$) and R$^{4+}$ ($\eta_R, \eta_R, 0$), order parameters have tetragonal $P4/mbm$ and orthorhombic $Imma$ symmetries, respectively. Specifically, for octahedral rotation: $P4/mbm$, basis = $\{(0, 0, 1), (1, 0, 0), (0, 1/2, 0)\}$ with Os ions at sites $2a$ and site symmetry $4/m$ on the b-axis, and, for octahedral tilting: $Imma$, basis = $\{(1, 0, 0), (0, 1, 0), (0, 0, 1)\}$ with Os ions at sites $4a$ and site symmetry $2/m$ on the $a$-axis, with basis vectors expressed in terms of an orthorhombic $Pnma$ cell.

Quadrupole contributions engaged in Templeton-Templeton scattering using an $E1 - E1$ event are,

$$\langle T^2_{+1}\rangle' \propto (xz) \text{ at } (h, 0, 0); \quad \langle T^2_{+1}\rangle'' \propto (yz) \text{ at } (h, h, 0); \quad \langle T^2_{+2}\rangle'' \propto (xy) \text{ at } (0, k, 0). \quad \text{(C1)}$$

The quadrupole for octahedral rotation using P4/mbm is provided in the main text in equation (B1). The corresponding quadrupole for Imma with rotation symmetry 2/m on the a-axis is,

$$\begin{pmatrix} p & 0 & 0 \\ 0 & q & r \\ 0 & r & s \end{pmatrix}, \quad \text{(C2)}$$

with trace (p + q + s) = 0. Rotations using angles $\theta_o$ & $\varphi_o$ defined in (B2) preserve the condition on the trace, and the element (yz) = r when rotation angles are set to zero. Complete results are,

$$\langle T^2_{+2}\rangle'' \propto (xy) = (1/2)\sin\theta_o[(q-s)\sin(2\varphi_o) + 2r\cos(2\varphi_o)],$$

$$\langle T^2_{+1}\rangle' \propto (xz) = (1/2)\sin(2\theta_o)[-p + q\sin^2\varphi_o + s\cos^2\varphi_o + r\sin(2\varphi_o)],$$

$$\langle T^2_{+1}\rangle'' \propto (yz) = (1/2)\cos\theta_o[(q-s)\sin(2\varphi_o) + 2r\cos(2\varphi_o)]. \quad \text{(C3)}$$

Dependence on $\theta_o$ and $\varphi_o$ in these expressions factorizes, because the two angles are associated with two distinct order parameters.

*valerio.scagnoli@psi.ch